\documentclass[amssymb,aps,floatfix,floats,preprintnumbers,showpacs,showkeys,twocolumn]{revtex4}
\usepackage{rcs}
\usepackage{color,graphicx}
\usepackage{dcolumn}
\usepackage{bm}
\usepackage[normalem]{ulem}
\usepackage{chngcntr}
\usepackage[thinspace,thinqspace]{SIunits}
\usepackage{natbib}

\newcommand{\et}{\textit{et al.}}
\newcommand{\muB}{$\mu_{\textrm{B}}$}

\newcommand{\YRS}{YbRh$_{2}$Si$_{2}$}
\newcommand{\YRSC}{Yb(Rh$_{1-x}$Co$_{x}$)$_{2}$Si$_{2}$}
\newcommand{\YRSCa}{Yb(Rh$_{0.73}$Co$_{0.27}$)$_{2}$Si$_{2}$}
\newcommand{\YRSCb}{Yb(Rh$_{0.79}$Co$_{0.21}$)$_{2}$Si$_{2}$}
\newcommand{\YRSCc}{Yb(Rh$_{0.82}$Co$_{0.18}$)$_{2}$Si$_{2}$}
\newcommand{\YRSCd}{Yb(Rh$_{0.88}$Co$_{0.12}$)$_{2}$Si$_{2}$}

\newcommand{\TN}{$T_{\textrm{N}}$}
\newcommand{\TL}{$T_{\textrm{L}}$}
\newcommand{\BN}{$B_{\textrm{N}}$}
\newcommand{\BL}{$B_{\textrm{L}}$}
\begin{document}
\preprint{APS/123-QED}
\title{Evolution from Ferromagnetism to Antiferromagnetism in \YRSC}
\author{S.~Hamann$^{1}$, J.~Zhang$^{1,2}$, D.~Jang$^{1}$, A. Hannaske$^{1}$, L. Steinke$^{1,3}$, S.~Lausberg$^{1}$, L.~Pedrero$^{1}$, C.~Klingner$^{1}$, M. Baenitz$^{1}$, F.~Steglich$^{1,2,4}$, C.~Krellner$^{1,5}$, C.~Geibel$^{1}$ and M.~Brando$^{1}$}
\affiliation{$^{1}$Max Planck Institute for Chemical Physics of Solids, D-01187 Dresden, Germany\\
$^{2}$Center of Correlated Matter, Zheijiang University, CHN-310058 Hangzhou, China\\
$^{3}$Department of Physics, Texas A\&M University, College Station, Texas 77843-4242, USA\\
$^{4}$Institute of Physics, Chinese Academy of Sciences, Beijing 100190, China\\
$^{5}$Institute of Physics, Goethe University Frankfurt, D-60438 Frankfurt am Main, Germany}
\date{\today}
\begin{abstract}
\YRSC\ is a model system to address two challenging problems in the field of strongly correlated electron systems: The first is the intriguing competition between ferromagnetic (FM) and antiferromagnetic (AFM) order when approaching a magnetic quantum critical point (QCP). The second is the occurrence of magnetic order along a very hard crystalline electric field (CEF) direction, i.e. along the one with the smallest available magnetic moment. Here, we present a detailed study of the evolution of the magnetic order in this system from a FM state with moments along the very hard $c$ direction at $x = 0.27$ towards the yet unknown magnetic state at $x = 0$. We first observe a transition towards an AFM canted state with decreasing $x$ and then to a pure AFM state. This confirms that the QCP in \YRS\ is AFM, but the phase diagram is very similar to those observed in some inherently FM systems like NbFe$_{2}$ and CeRuPO, which suggests that the basic underlying instability might be FM. Despite the huge CEF anisotropy the ordered moment retains a component along the $c$-axis also in the AFM state. The huge CEF anisotropy in \YRSC\ excludes that this hard-axis ordering originates from a competing exchange anisotropy as often proposed for other heavy-fermion systems. Instead, it points to an order-by-disorder based mechanism.
\end{abstract}
\maketitle
A comprehensive understanding of magnetic quantum phase transitions (QPTs) and associated quantum critical points (QCPs) is considered to be a fundamental step in attempting to reveal the physics of strongly correlated electrons. Despite more than 40 years of research, there are still QPTs, observed in particular in exotic metals, that are far from been understood~\cite{Gegenwart2008,Loehneysen2007,Brando2016}. This is mainly due to the complexity of these systems, the properties of which are often governed by magnetic anisotropies, competing interactions, geometric frustration, Fermi surface instabilities etc., i.e. by not just one, but multiple energy scales. This, on the other hand, results in the appearance of fascinating states of matter near QCPs, as, e.g., spin liquids~\cite{Friedemann2009}.

In this respect, a prototypical and well studied example is the tetragonal Kondo lattice \YRS~\cite{Trovarelli2000}. Despite a large Kondo temperature $T_{\textrm{K}} \approx 25$\,K, this compound shows antiferromagnetic (AFM) order at \TN\ = 0.07\,K that can be suppressed either by a magnetic field or negative chemical pressure to reveal an intriguing QCP~\cite{Custers2003} whose nature is still strongly debated~\cite{Paschen2004,Friedemann2009,Friedemann2010,Kummer2015,Woelfle2015}. A detailed study of the magnetic fluctuations at this QCP is hindered by the lack of knowledge of the AFM ordered structure which is due to the very low \TN\ and the small ordered moment ($10^{-3}$\muB/Yb)~\cite{Ishida2003}. First attempts with inelastic neutron scattering have detected ferromagnetic (FM) fluctuations at low temperatures that evolve on cooling into incommensurate correlations located at $q = (0.14, 0.14, 0)$ just above \TN~\cite{Stock2012}. This agrees with previous experiments which indicate a large value of the in-plane susceptibility ($9 {\times} 10^{-6}$m$^{3}$/mol $\approx$ 0.18\,SI) and of the Sommerfeld-Wilson ratio ($\approx 30$), implying the presence of strong FM fluctuations~\cite{Gegenwart2005}. 

Although the AFM structure below \TN\ is unknown, the large crystalline electric field (CEF) anisotropy, with very different $g$-factors ($g_{c} \approx 0.2$ and $g_{ab} \approx 3.6$) along the $c$-axis and within the $ab$-plane~\cite{Gruner2012}, points to moments oriented mainly within the basal plane. Such anisotropy is seen in the uniform magnetic susceptibility which is much larger for fields applied in the basal plane as compared to fields along the $c$-axis. Also the fields needed to suppress the AFM state are strongly anisotropic, i.e., \BN($\perp$\textit{c}) = 0.06\,T and \BN($\parallel$\textit{c}) = 0.66\,T~\cite{Gegenwart2002}. 

\begin{figure}[t]
	\centering
	\includegraphics[width=\columnwidth]{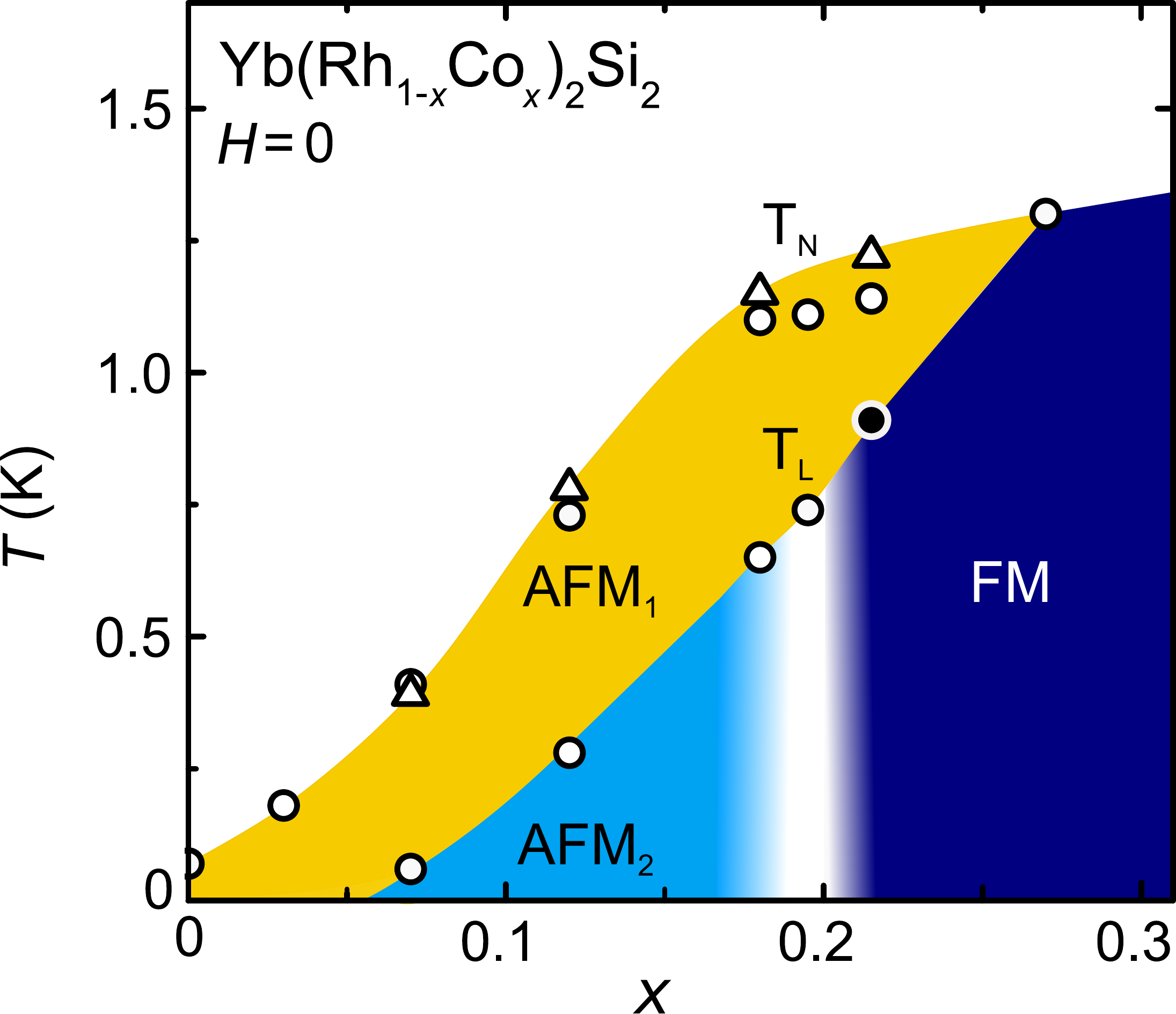}
	\caption{Phase diagram of \YRSC. The ferromagnetic (FM) phase is separated from two antiferromagnetic phases, AFM$_{1}$ and AFM$_{2}$, by first order lines. The circles are taken from Refs.~\onlinecite{Klingner2011} and~\onlinecite{Westerkamp2008}. The triangles corresponds to the peaks in $\chi'(T)$ of this work. The filled point indicates a first order transition. Since we have investigated only samples with \textit{x} = 0.18 and 0.21, we do not know the exact location of the line between the FM and the AFM$_{2}$ phases and we left this area uncolored.}
	\label{fig1}
\end{figure}

In order to have better access to the AFM state, it is convenient to enhance \TN\ and the size of the ordered moment. This was done by applying hydrostatic pressure~\cite{Mederle2001,Knebel2006} which stabilizes the magnetic Yb$^{3+}$ state or by substituting the isoelectronic smaller Co for Rh: In fact, the whole series \YRSC\ crystallizes in the same ThCr$_{2}$Si$_{2}$ structure~\cite{Klingner2011}. Increasing $x$ has a strong effect on the relevant energy scales: i) the Kondo temperature decreases causing an enhancement of \TN, ii) the CEF anisotropy becomes weaker and iii) FM correlations increase~\cite{Klingner2011}. In addition, a second phase transition at \TL\ $<$ \TN\ occurs~\cite{Westerkamp2008,Klingner2011}. The nature of the phase below \TL\ was believed to be AFM until it was discovered that \YRSCa\ displays FM order below \TN\ = \TL\ = 1.3\,K~\cite{Lausberg2013} (cf. Fig.\ref{fig1}). It is worth noting that FM ordering was previously proposed to occur under hydrostatic pressure by Knebel \et~\cite{Knebel2006}.

This discovery immediately raised the question about a possible FM ordering in \YRS, which might have been overseen because of the extremely small ordered moment. The presence of a FM (instead of an AFM) QCP would have a strong impact in the field, since \YRS\ is one of the few pivotal systems considered for the development of contemporary theories of AFM QCPs. It is therefore essential to determine the nature of the magnetic order for $0 \leq x \leq 0.27$. In this Letter we show that \YRSC\ evolves, with decreasing $x$, from a FM ground state at $x = 0.27$ to a canted AFM and then to a pure AFM ground state (AFM$_{1}$ and AFM$_{2}$ in Fig.~\ref{fig1}, respectively). Thus, eventually the QCP in \YRS\ is of AFM nature, but a comparison with the phase diagrams of other materials close to a FM instability~\cite{Brando2016}, like NbFe$_{2}$~\cite{Brando2008}, CeRuPO~\cite{Kotegawa2013} or PrPtAl~\cite{Jabbar2015}, suggests that the dominant incipient instability might be the FM one.

The biggest surprise about the discovery of FM ordering in \YRSCa\ is the fact that the ordered moments are aligned along the $c$-axis, despite the moment provided by the CEF ground state is six times smaller along the $c$ direction than that in the basal plane~\cite{Gruner2012,Lausberg2013,Andrade2014}. This is completely unexpected and cannot be understood within standard theories of magnetism, since the gain in energy in the ordered state is expected to be proportional to the square of the size of the ordered moment. Remarkably, it has recently been realized that ordering along the hard CEF axis is quite common in ferromagnetic Kondo systems~\cite{Bonville1992,Krellner2008,Araki2003,Krueger2014,Hafner2018} (cf. section E of the Supplementary Material(SM)). A reorientation of the moment from the easy to the hard CEF direction has also been reported for a few AFM Kondo lattices with increasing hybridization strength and approaching the QCP~\cite{Kondo2013,Khalyavin2014,Kobayashi2014}. Therefore, the ordering along the hard CEF direction seems to be a common feature in Kondo systems, especially in FM ones, which is yet not understood and thus deserves a dedicated study. Analyzing our data, we realized that \YRSC\ is a key system to address this problem. We found that \YRSC\ retains a component of the ordered moment along the $c$ direction with decreasing $x$, likely until $x = 0$. The huge anisotropy at low $x$ definitely excludes that the mechanism proposed for the AFM systems is valid here. Instead, we suggest that the origin of this hard-axis ordering is an order-by-disorder mechanism.

We present first in Fig.\ref{fig1} the main result of our work, i.e., the zero field phase diagram of \YRSC\ with $0 \leq x \leq 0.3$ and then show how it was constructed. This phase diagram consists mainly of four phases: a paramagnetic phase (PM), a FM phase and two AFM phases, AFM$_{1}$ for \TL\ $< T  \leq$ \TN\ and AFM$_{2}$ for $0 < T \leq$ \TL. In the FM phase the moments are aligned mainly along the $c$-axis as described in Ref.~\cite{Lausberg2013}. In the AFM$_{1}$ phase the propagation vector has a component within the $ab$-plane, and a component of the ordered moment is along the $c$-axis. The latter does not change between the AFM$_{1}$ and AFM$_{2}$ phases.

\begin{figure}[t]
	\includegraphics[width=\columnwidth]{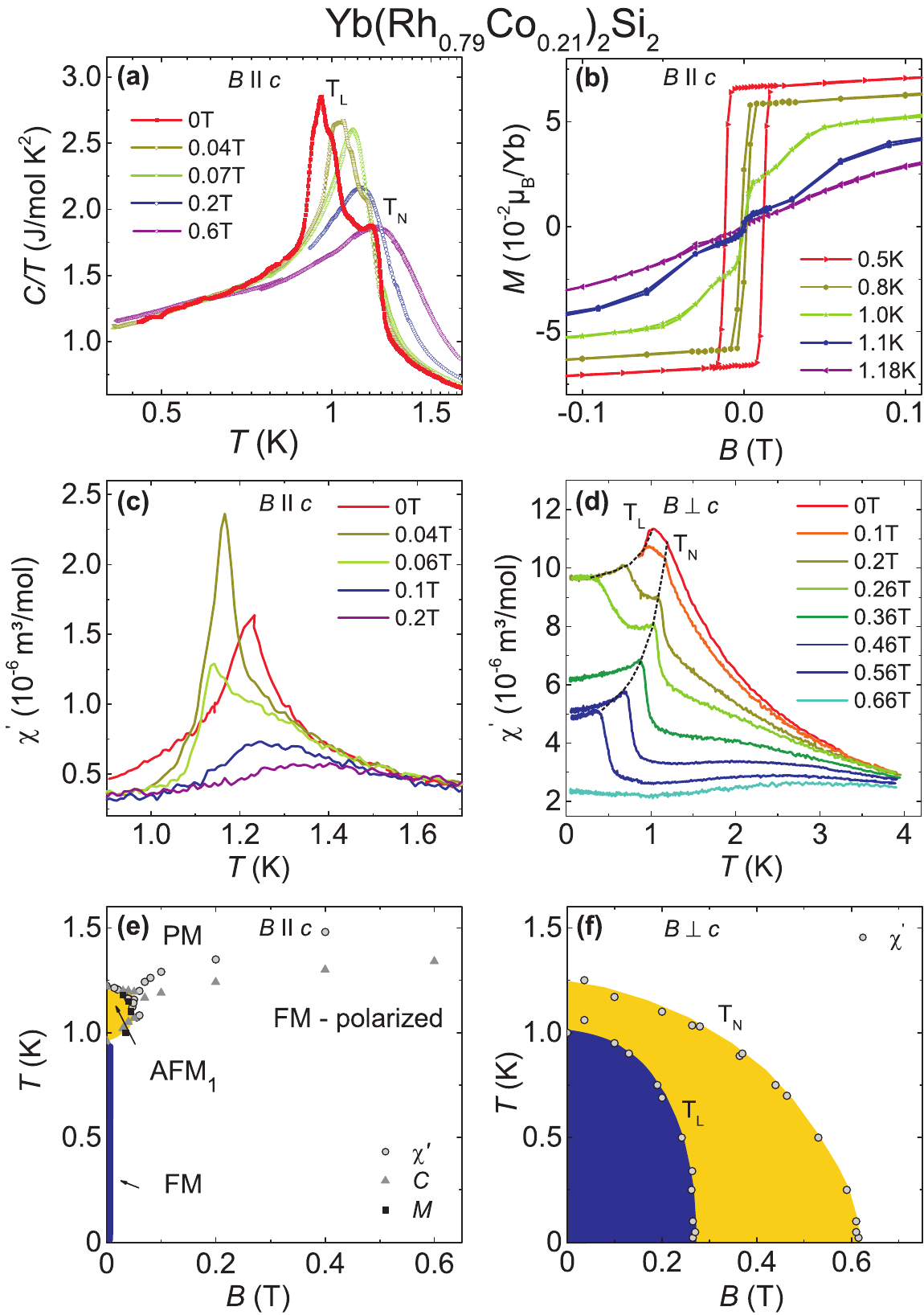}
	\caption{Selection of measurements on \YRSCb\ with \textit{B}$\parallel$\textit{c} and \textit{B}$\perp$\textit{c}. (a) Specific heat $C(T)$, (b) magnetization $M(B)$, (c,d) ac susceptibility $\chi'(T)$ and the phase diagrams (e) with \textit{B}$\parallel$\textit{c} and (f) with \textit{B}$\perp$\textit{c}.}
	\label{fig2}
\end{figure}

In the following, we present selected data for samples with \textit{x} = 0.21, 0.18 and 0.12 from which we constructed the phase diagram. Fig.\ref{fig2} shows several measurements performed on \YRSCb\ with \textit{B}$\parallel$\textit{c} to look for a FM response along the $c$-axis and one measurement with \textit{B}$\perp$\textit{c} for comparison. At \textit{B} = 0 we observe a large peak in the specific heat at \TL$ \approx $0.95\,K and a broad shoulder at \TN = 1.2\,K (cf. red curve in Fig.\ref{fig2}a) which correspond to a transition of first order at \TL\ and a mean-field-like transition at \TN, in agreement with Ref.~\cite{Klingner2011}. With increasing field the transition at \TL\ shifts to higher temperatures as expected for a FM order. The opposite is observed for \textit{B}$\perp$\textit{c}~\cite{Klingner2009}. Magnetization $M(B)$ with \textit{B}$\parallel$\textit{c} is displayed in Fig.\ref{fig2}b: Right below \TN, $M(B)$ shows a very small remanent magnetization of about 0.01\,\muB\ with a tiny hysteresis and a metamagnetic transition at \BN$ \approx 0.03$\,T pointing to a canting of the moments. With decreasing $T$, the remanent magnetization and the hysteresis loop increase while \BN\ decreases until, below \TL, we have a pure FM hysteresis, as seen for \textit{x} = 0.27~\cite{Lausberg2013}. 
The remanent magnetization of 0.06\,\muB\ along the $c$-axis is half of that measured in the sample with \textit{x} = 0.27. This reflects the higher CEF anisotropy and $T_{\textrm{K}}$ in \YRSCb\ compared to those in \YRSCa. The magnetic anisotropy is also reflected in the behavior of the ac susceptibility $\chi'(T)$ shown in Figs.\ref{fig2}c and \ref{fig2}d for \textit{B}$\parallel$\textit{c} and \textit{B}$\perp$\textit{c}, respectively. For \textit{B}$\parallel$\textit{c}, $\chi'(T)$ detects the transition at \TN\ = 1.22\,K in form of a broad peak but misses that at \TL. This is because the modulation field of $\chi'(T)$ is smaller than the coercive field for $T <$ \TN, and below \TL\ the coercive field becomes even larger. On the other hand, for \textit{B}$\perp$\textit{c}, $\chi'(T)$ detects both transitions in form of a kink and a drop at \TN\ and \TL, respectively (dashed lines in Fig.\ref{fig2}d). The peak for \textit{B}$\parallel$\textit{c} becomes higher and sharper at \textit{B} = 0.04\,T with a signature in $\chi''(T)$ (not shown) indicating dissipation~\cite{Hamann2018}. $\chi'(T)$ reaches a value of $2.4 {\times} 10^{-6}$\,m$^{3}$/mol which is about four times smaller than that measured with \textit{B}$\perp$\textit{c} (see Fig.\ref{fig2}d). For $B {>} 0.04$\,T, $\chi'(T)$ broadens and loses intensity. The phase diagrams in both field directions are shown in Figs.\ref{fig2}e,f: The AFM$_{1}$ phase covers a very small area in the \textit{B-T} phase diagram for \textit{B}$\parallel$\textit{c}. For \textit{B}$\perp$\textit{c} both the FM and the AFM$_{1}$ phases are suppressed at finite fields of 0.26\,T and 0.6\,T, respectively. Therefore, \YRSCb\ goes from a canted AFM state into a FM ground state with moments along the $c$-axis, through a transition which is first order at \textit{B} = 0. Similar magnetic phase diagrams have been observed in other materials, like NbFe$_{2}$~\cite{Friedemann2018} and can be reproduced by theories which consider two order parameters, one for the AFM phase and the other one for the FM phase~\cite{Moriya1977}.    
\begin{figure}
\centering
\includegraphics[width=\columnwidth]{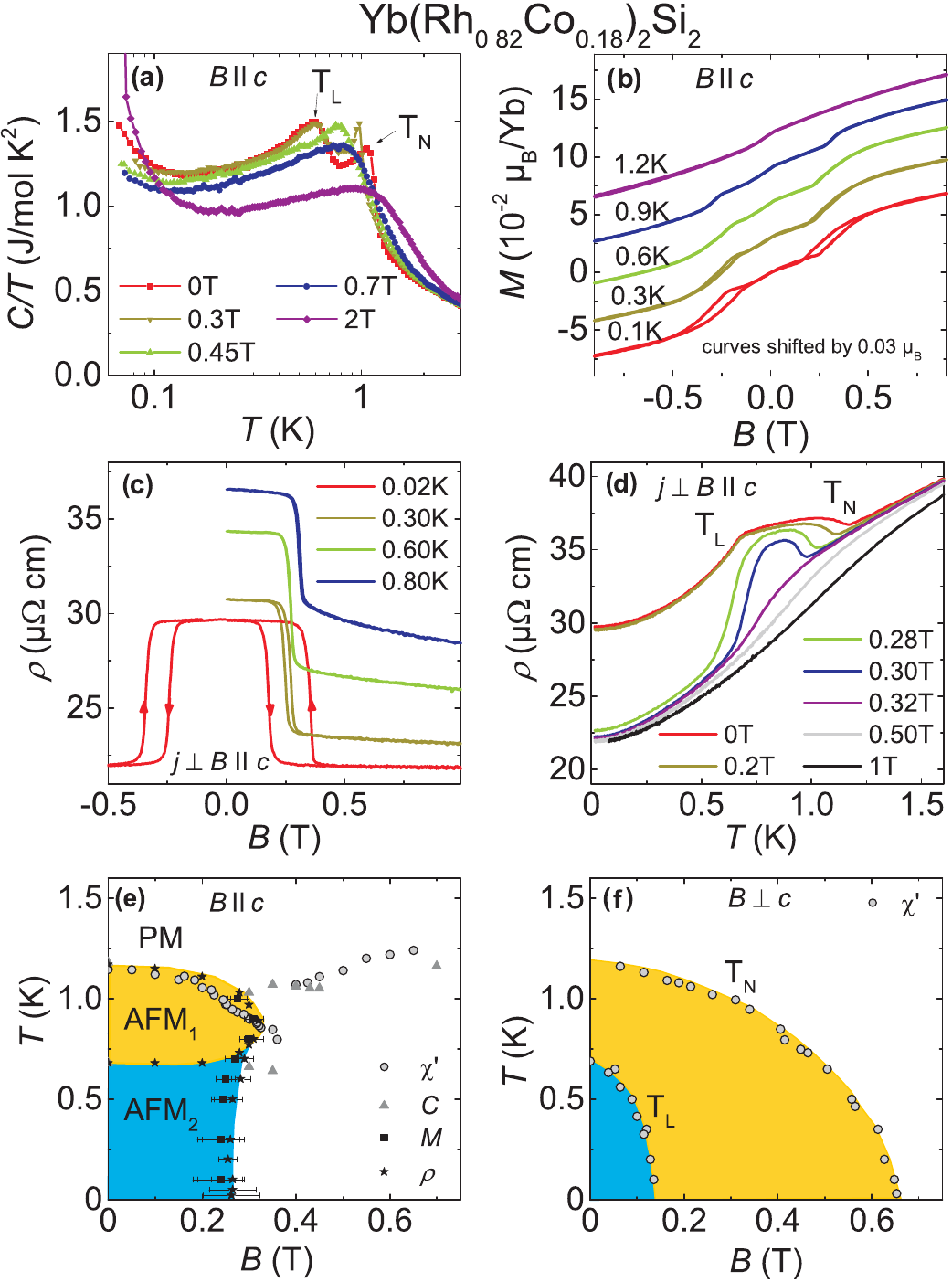}
\caption{Selection of measurements on \YRSCc\ with \textit{B}$\parallel$\textit{c}. (a) Specific heat $C(T)$, (b) magnetization $M(B)$, (c,d) resistivity $\rho(T,B)$ and the phase diagrams (e) with \textit{B}$\parallel$\textit{c} and (f) with \textit{B}$\perp$\textit{c}.}
\label{fig3}
\end{figure}

We discuss now next sample, \YRSCc. Fig.\ref{fig3} shows selected measurements with \textit{B}$\parallel$\textit{c}. The specific heat detects two second order phase transitions at \TN\ = 1.1\,K and \TL\ = 0.65\,K (Fig.\ref{fig3}a)~\cite{Klingner2011}. Interestingly, with increasing \textit{B}$\parallel$\textit{c}, \TN\ decreases slightly but \TL\ does not change. Above $B \approx 0.35$\,T both signatures join into a common broad peak that shifts to high $T$ with increasing $B$. The magnetization is shown in Fig.\ref{fig3}b. There is no evident remanent magnetization along the $c$-axis. Below \TN, $M(B)$ isotherms develop metamagnetic jumps at a finite critical field $B_{\textrm{N}} \approx 0.25$\,T. The jump at \BN\ is substantial, $\Delta M(B_{\textrm{N}})$ = 0.02\muB, considering that the magnetization above \BN\ is 0.05\muB, i.e., very close to the saturation value. Decreasing $T$ below \TL\ does not affect the shape of the isotherms, but for $T \leq 0.3$\,K the jumps become hysteretic indicating a sort of spin-flop first order transition. On the other hand, magnetization isotherms for $T <$ \TL\ with \textit{B}$\perp$\textit{c} show first a weak metamagnetic-like transition at \BL\ $\approx  0.13$\,T (see Fig.1a in the SM~\cite{Sakakibara1994,Pedrero2013}) and a kink at \BN\ $\approx 0.7$\,T with no remanent magnetization nor hysteresis~\cite{Pedrero2013}. The signatures in $\chi'(T)$ for both field directions are similar to those seen in \YRSCb. Magnetoresistance measurements with current \textit{j}$\perp$\textit{c} and \textit{B}$\parallel$\textit{c} (Fig.\ref{fig3}c), show a large hysteresis at \BN\ (the asymmetry is due to the remanent field of the 20\,T magnet), confirming the first order nature of the spin-flop transition. Interestingly, the $T$-dependence of $\rho(T)$ shows a clear jump at \TN\ (Fig.\ref{fig3}d) indicating the opening of a gap at the Fermi level and implying that the propagation vector in the AFM$_{1}$ phase has a component within the $ab$-plane resulting in a sizeable gap in the plane. 
\begin{figure}
\centering
\includegraphics[width=\columnwidth]{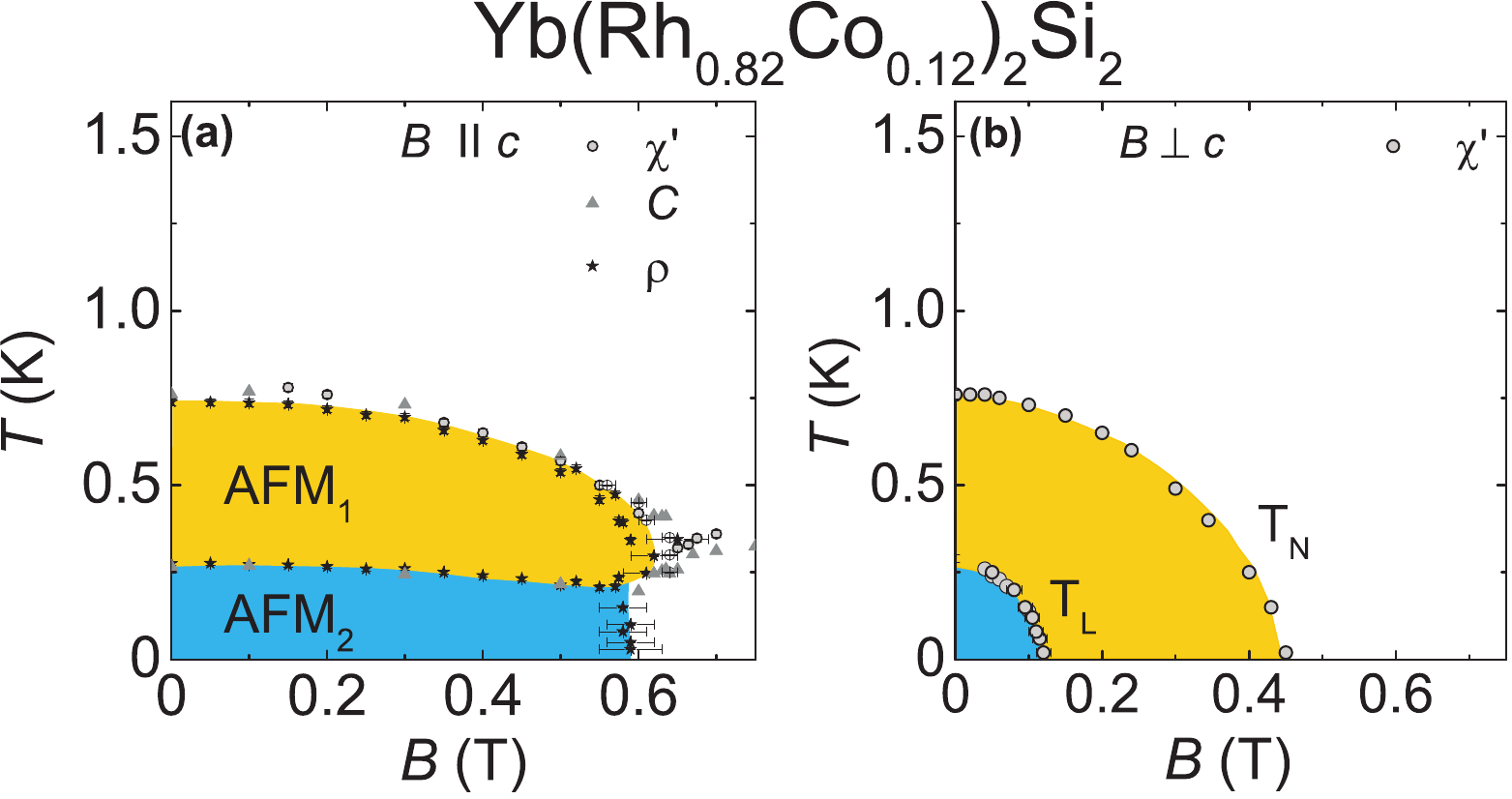}
\caption{Phase diagrams of \YRSCd\ with (a) \textit{B}$\parallel$\textit{c} and (b)  \textit{B}$\perp$\textit{c}.}
\label{fig4}
\end{figure}

All measurements leave us with the phase diagrams shown in Figs.\ref{fig3}e,f. \YRSCc\ shows an AFM state below \TN, AFM$_{1}$, in which we could not detect sizeble canting (i.e., remanent magnetization along $c$) and a second AFM state, AFM$_{2}$, below \TL. Both AFM states can be suppressed by magnetic fields \textit{B}$\parallel$\textit{c} and \textit{B}$\perp$\textit{c}, but the phase transition for \textit{B}$\parallel$\textit{c} is first order and ends at a multicritical point (MCP) located at about 0.9\,K and 0.3\,T. This indicates that the moments in the AFM$_{2}$ phase have a component along the $c$-axis which flips at the critical field. Furthermore, the fact that the phase line at \TL\ is horizontal (Fig.~\ref{fig3}e) indicates that $dT/dB = 0$ at \TL, which implies (Ehrenfest equation) that $\left( \partial M_{1}/\partial B \right)_{T} = \left( \partial M_{2}/\partial B \right)_{T}$, where $M_{1}$ and $M_{2}$ are the magnetizations in the phase AFM1 and AFM2, respectively. This can be also verified by looking at the slope $dM/dB$ of the isothermal magnetization in Fig.~\ref{fig3}b, which does not change across \TL. This implies that the evolution of the component $\parallel c$ of the ordered moment is linear in $T$ across \TL, i.e. it is not affected by \TL, and only the evolution of the component $\perp c$ changes. That explains why the transition at \TL\ can not be clearly seen in $\chi'(T) = dM/dB$ with \textit{B}$\parallel$\textit{c}. Thus, if the moments in the AFM$_{2}$ state have a component along the $c$-axis the shape of the phase diagram suggests that this component is also present in the AFM$_{1}$ phase. Also the analysis of the nuclear Schottky contribution to the specific heat, visible as a $T^{-3}$ increase in $C(T)/T$ below 0.2\,K (cf. Fig.\ref{fig3}a and section C of the SM~\cite{Wilhelm2004,Steppke2010}) provides another indication that in the AFM$_{2}$ phase the moments have a component along the $c$-axis which flips at the critical field.

To see whether both AFM phases extend to lower \textit{x}, we take a look at the phase diagrams of the next sample with \textit{x} = 0.12 shown in Fig.\ref{fig4}. It is very similar to that of the sample with \textit{x} = 0.18 but with a larger AFM$_{1}$ area of the \textit{B-T} phase diagram with \textit{B}$\parallel$\textit{c}. All measurements done on \YRSCd\ are very similar to those done on \YRSCc\ (see, e.g., $\rho(B)$ in the SM), but the features are weaker due to the smaller ordered moment. This signifies that the same AFM$_{1}$ and AFM$_{2}$ phases of \YRSCc\ are also present in \YRSCd.

To summarize our experimental results: The phase diagram of \YRSC\ can be drawn as in Fig.\ref{fig1} with a FM ground state for \textit{x} = 0.27 and 0.21 with moments along the $c$-axis and an AFM$_{2}$ ground state where the moments possess a component along the $c$-axis. This component does not change between the AFM$_{2}$ and AFM$_{1}$ phase, but becomes smaller with $x \rightarrow 0$. This is due to two effects: i) a decrease in size of the moments because of the increasing Kondo screening and ii) a rotation of the moments towards the $ab$-plane. The small canted moment observed for \textit{x} = 0.21 in the AFM$_{1}$ phase vanishes or is not detectable anymore for $x \leq 0.18$. This might be due to the very small component of the moments along \textit{c} or to a slight change in the spin structure. It might be therefore helpful to perform high-resolution polarized neutron scattering experiments on \YRS\ to look for a component of the moments along the $c$-axis.

We discuss now the implication of our results for the problem of hard CEF direction ordering in Kondo lattices and for the nature of the QCP in \YRS. Having a component of the ordered moment along the CEF $c$-axis even at smaller $x$ is very surprising, since the CEF anisotropy is even higher. For those AFM systems that show ordering along the hard axis, it has been proposed that this is due to a competing anisotropy (respective to the direction of the moment) of the exchange interaction which overcomes the CEF anisotropy~\cite{Kondo2013,Khalyavin2014,Kobayashi2014}. However, in our case the huge CEF anisotropy ($> 10$) in slightly Co-doped \YRS\ would require a huge inverse anisotropy in the exchange interactions, which seems very unlikely. Indeed, the homologues GdRh$_{2}$Si$_{2}$ and EuRh$_{2}$Si$_{2}$, which do not present CEF effects because their $4f$-electron moment which is a pure spin $S = 7/2$, show only a very weak magnetic anisotropy~\cite{Seiro2014,Kliemt2017}. This implies that in the RERh$_{2}$Si$_{2}$ series (RE: rare earth) the anisotropy of the exchange interaction respective to the orientation of the moment is rather weak. Thus, the mechanism which has so far been proposed for hard CEF direction ordering can safely be excluded for the present case. Instead, our results point to a different origin. Since the proximity of all these systems to a QCP results in large fluctuations, a mechanism based on an order-by-disorder process seems to be a better candidate. It is, e.g., conceivable that fluctuations of the large CEF in-plane moment stabilize an ordering along the hard CEF direction. Such a situation has been explicitly demonstrated in a two-band model by F. Kr\"uger \et~\cite{Krueger2014}. We propose that a such a mechanism is also responsible for the hard-axis ordering in \YRSC.

Since we did not find any evidence for a FM ordering for $x \leq 0.18$, our results indicate that the QCP in \YRS\ is AFM. However, we note that the phase diagram shown in Fig.~\ref{fig1} is very similar to those observed in prototypical FM systems in which the FM QCP is avoided by switching to an AFM state with a small propagation vector~\cite{Belitz1997}, like, e.g., NbFe$_{2}$~\cite{Brando2008}. In \YRS\ inelastic neutron scattering found FM-fluctuations which on cooling evolved to incommensurate correlations located at $q = (0.14, 0.14, 0)$~\cite{Stock2012}. This propagation vector is similar in size to $q = (0, 0, 0.157)$ observed in NbFe$_{2}$ in its AFM regime very close to the QCP~\cite{Niklowitz2017}. This similarity and the presence of a stable FM state for $0.21 \leq x \leq 0.5$ in \YRSC\ suggests that the basic underlying magnetic instability in \YRS\ might be FM, but that eventually, before reaching the QCP, an AFM state with a long modulation emerges as observed in other FM systems~\cite{Brando2016}. This is supported by NMR studies which indicated dominant FM correlations being overwhelmed by AFM ones only at very low $T$~\cite{Ishida2003}. On top of this scenario which is of general relevance for all metallic FM systems close to the QCP, the present study emphasizes a further feature which is likely specific to $4f$-Kondo lattices close to a FM QCP: The unexpected orientation of the ordered moment along the hard CEF direction. In fact, it has been found that almost all FM Kondo-lattice systems show ordering with moments along the CEF hard direction~\cite{Hafner2018}, as found in \YRSC.

Our results have a further important consequence: The phase boundary line between the AFM$_{2}$ phase and the PM phase is first order and terminates at a MCP at finite temperature (see, e.g., Fig.\ref{fig3}e). If this point was shifted to \textit{T} = 0 at a certain concentration between 0.12 and 0, than this point would have the nature of a field-induced quantum MCP in remarkable agreement with predictions of Misawa \et~\cite{Misawa2008,Misawa2009} and very similar to what has been observed in NbFe$_{2}$~\cite{Friedemann2018}.

\begin{acknowledgments}
We are indebted to T. L\"uhmann, A. Steppke, O. Stockert and S. Wirth for useful discussions as well as to C. Klausnitzer for experimental support. Part of the work was funded by the Deutsche Forschungsgemeinschaft (DFG) Research Unit 960 'Quantum Phase Transitions' and the DFG projects BR 4110/1-1 and KR 3831/4-1.
\end{acknowledgments}
\bibliography{hamann_prl}

\begin{thebibliography}{10}

\bibitem{Gegenwart2008}
P.~Gegenwart, Q.~Si, and F.~Steglich,
\newblock Nature Phys. {\bf 4}, 186 (2008).

\bibitem{Loehneysen2007}
H.~v. L\"ohneysen, A.~Rosch, M.~Vojta, and P.~W\"olfle,
\newblock Rev. Mod. Phys. {\bf 79}, 1015 (2007).

\bibitem{Brando2016}
M.~Brando, D.~Belitz, F.~M. Grosche, and T.~R. Kirkpatrick,
\newblock Rev. Mod. Phys. {\bf 88}, 025006 (2016).

\bibitem{Friedemann2009}
S.~Friedemann {\em et~al.},
\newblock Nature Phys. {\bf 5}, 465 (2009).

\bibitem{Trovarelli2000}
O.~Trovarelli {\em et~al.},
\newblock Phys. Rev. Lett. {\bf 85}, 626 (2000).

\bibitem{Custers2003}
J.~Custers {\em et~al.},
\newblock Nature {\bf 424}, 524 (2003).

\bibitem{Paschen2004}
S.~Paschen {\em et~al.},
\newblock Nature {\bf 432}, 881 (2004).

\bibitem{Friedemann2010}
S.~Friedemann {\em et~al.},
\newblock Proc. Natl. Acad. Sci. {\bf 107}, 14547 (2010).

\bibitem{Kummer2015}
K.~Kummer {\em et~al.},
\newblock Phys. Rev. X {\bf 5}, 011028 (2015).

\bibitem{Woelfle2015}
P.~W\"olfle and E.~Abrahams,
\newblock Phys. Rev. B {\bf 92}, 155111 (2015).

\bibitem{Ishida2003}
K.~Ishida {\em et~al.},
\newblock Phys. Rev. B {\bf 68}, 184401 (2003).

\bibitem{Stock2012}
C.~Stock {\em et~al.},
\newblock Phys. Rev. Lett. {\bf 109}, 127201 (2012).

\bibitem{Gegenwart2005}
P.~Gegenwart, J.~Custers, Y.~Tokiwa, C.~Geibel, and F.~Steglich,
\newblock Phys. Rev. Lett. {\bf 94}, 076402 (2005).

\bibitem{Gruner2012}
T.~Gruner {\em et~al.},
\newblock Phys. Rev. B {\bf 85}, 035119 (2012).

\bibitem{Gegenwart2002}
P.~Gegenwart {\em et~al.},
\newblock Phys. Rev. Lett. {\bf 89}, 056402 (2002).

\bibitem{Mederle2001}
S.~Mederle {\em et~al.},
\newblock J. Magn. Magn. Mater. {\bf 226--230}, 254 (2001).

\bibitem{Knebel2006}
G.~Knebel {\em et~al.},
\newblock J. Phys. Soc. Jpn. {\bf 75}, 114709 (2006).

\bibitem{Klingner2011}
C.~Klingner {\em et~al.},
\newblock Phys. Rev. B {\bf 83}, 144405 (2011).

\bibitem{Westerkamp2008}
T.~Westerkamp, P.~Gegenwart, C.~Krellner, C.~Geibel, and F.~Steglich,
\newblock Physica B: Cond. Matt. {\bf 403}, 1236 (2008).

\bibitem{Lausberg2013}
S.~Lausberg {\em et~al.},
\newblock Phys. Rev. Lett. {\bf 110}, 256402 (2013).

\bibitem{Brando2008}
M.~Brando {\em et~al.},
\newblock Phys. Rev. Lett. {\bf 101}, 026401 (2008).

\bibitem{Kotegawa2013}
H.~Kotegawa {\em et~al.},
\newblock J. Phys. Soc. Jpn. {\bf 82}, 123711 (2013).

\bibitem{Jabbar2015}
G.~Abdul-Jabbar {\em et~al.},
\newblock Nat. Phys. {\bf 11}, 321 (2015).

\bibitem{Andrade2014}
E.~C. Andrade, M.~Brando, C.~Geibel, and M.~Vojta,
\newblock Phys. Rev. B {\bf 90}, 075138 (2014).

\bibitem{Bonville1992}
P.~Bonville {\em et~al.},
\newblock Physica B {\bf 182}, 105 (1992).

\bibitem{Krellner2008}
C.~Krellner and C.~Geibel,
\newblock J. Crystal Growth {\bf 310}, 1875 (2008).

\bibitem{Araki2003}
S.~Araki {\em et~al.},
\newblock Phys. Rev. B {\bf 68}, 024408 (2003).

\bibitem{Krueger2014}
F.~Kr\"uger, C.~J. Pedder, and A.~G. Green,
\newblock Phys. Rev. Lett. {\bf 113}, 147001 (2014).

\bibitem{Hafner2018}
D.~Hafner {\em et~al.},
\newblock arXiv:1901.04288 .

\bibitem{Kondo2013}
A.~Kondo {\em et~al.},
\newblock J. Phys. Soc. Jpn. {\bf 82}, 054709 (2013).

\bibitem{Khalyavin2014}
D.~D. Khalyavin {\em et~al.},
\newblock Phys. Rev. B {\bf 89}, 064422 (2014).

\bibitem{Kobayashi2014}
R.~Kobayashi {\em et~al.},
\newblock J. Phys. Soc. Jpn. {\bf 83}, 104707 (2014).

\bibitem{Klingner2009}
C.~Klingner,
\newblock Diplom thesis, University of Dresden, 2009.

\bibitem{Hamann2018}
S.~Hamann,
\newblock {Ph.D.} thesis, University of Dresden, 2018.

\bibitem{Friedemann2018}
S.~Friedemann {\em et~al.},
\newblock Nature Phys. {\bf 14}, 62 (2018).

\bibitem{Moriya1977}
T.~Moriya and K.~Usami,
\newblock Solid State Commun. {\bf 23}, 935 (1977).

\bibitem{Sakakibara1994}
T.~Sakakibara, H.~Mitamura, T.~Tayama, and H.~Amitsuka,
\newblock Jpn. J. Appl. Phys. {\bf 33}, 5067 (1994).

\bibitem{Pedrero2013}
L.~Pedrero,
\newblock {Ph.D.} thesis, University of Dresden, 2013.

\bibitem{Wilhelm2004}
H.~Wilhelm, T.~L\"uhmann, T.~Rus, and F.~Steglich,
\newblock Rev. Sci. Instrum. {\bf 75}, 2700 (2004).

\bibitem{Steppke2010}
A.~Steppke {\em et~al.},
\newblock Phys. Status Solidi (b) {\bf 247}, 737 (2010).

\bibitem{Seiro2014}
S.~Seiro and C.~Geibel,
\newblock J. Phys.: Condens. Matter {\bf 26}, 046002 (2014).

\bibitem{Kliemt2017}
K.~Kliemt {\em et~al.},
\newblock Phys. Rev. B {\bf 95}, 134403 (2017).

\bibitem{Belitz1997}
D.~Belitz, T.~R. Kirkpatrick, and T.~Vojta,
\newblock Phys. Rev. B {\bf 55}, 9452 (1997).

\bibitem{Niklowitz2017}
P.~G. Niklowitz {\em et~al.},
\newblock arXiv:1704.08379 .

\bibitem{Misawa2008}
T.~Misawa, Y.~Yamaji, and M.~Imada,
\newblock J. Phys. Soc.Jpn. {\bf 77}, 093712 (2008).

\bibitem{Misawa2009}
T.~Misawa, Y.~Yamaji, and M.~Imada,
\newblock J. Phys. Soc. Jpn. {\bf 78}, 084707 (2009).

\end{thebibliography}
\bibliographystyle{h-physrev}
\end{document}